\documentclass[twocolumn]{jpsj2} 
\usepackage{bm,xspace}

\title{New Magnetic State in Incommensurate Magnetic Phase of Heavy-Fermion Superconductor CeRh$_{\bm{0.6}}$Co$_{\bm{0.4}}$In$_{\bm{5}}$}

\author{Makoto \textsc{Yokoyama}\thanks{E-mail address: makotti@mx.ibaraki.ac.jp}, Hiroshi \textsc{Amitsuka}$^{1}$, Kei \textsc{Matsuda}$^{1}$, Akifumi \textsc{Gawase}$^{1}$,\\ Narumi \textsc{Oyama}, Ikuto \textsc{Kawasaki}$^{1}$, Kenichi \textsc{Tenya}$^{1}$ and Hideki \textsc{Yoshizawa}$^{2}$}

\inst{
Faculty of Science, Ibaraki University, Mito 310-8512\\
$^{1}$Graduate School of Science, Hokkaido University, Sapporo 060-0810\\
$^{2}$Neutron Science Laboratory, Institute for Solid State Physics, The University of Tokyo, Tokai 319-1106\\
}

\def\runauthor{M.\ \textsc{Yokoyama} {\it et al.}}

\abst{
We performed elastic neutron scattering experiments on solid solution CeRh$_{1-x}$Co$_{x}$In$_5$ with $x=0.4$ to clarify the nature of the antiferromagnetic (AF) phase in the vicinity of the quantum critical point. We observed the incommensurate AF order below $T_{\rm Nh}=3.5\ {\rm K}$. The structure of the incommensurate AF order is basically unchanged from that for pure CeRhIn$_5$. We further found the evolution of a new commensurate AF order with the modulation of $q_{\rm c}=(1/2,1/2,1/2)$ below $T_{\rm Nc}=2.9\ {\rm K}$. The volume-averaged moments for the incommensurate and commensurate AF phases are $0.38\ \mu_{\rm B}/{\rm Ce}$ and $0.21\ \mu_{\rm B}/{\rm Ce}$, respectively, which are reduced from the incommensurate AF moment ($0.75\ \mu_{\rm B}/{\rm Ce}$) for pure CeRhIn$_5$. We suggest from these results that the commensurate magnetic correlation may be tightly coupled with the superconductivity observed below $T_{\rm c}=1.4\ {\rm K}$ 
}

\kword{heavy-fermion system, superconductivity, magnetism, neutron scattering, CeRhIn$_{\bm 5}$, CeCoIn$_{\bm 5}$}

\begin{document}
\maketitle
In recent years, there has been growing interest in the nature of the unusual magnetic fluctuations and new ordered phases observed near the quantum critical point (QCP) in heavy-fermion systems. The QCP, which is associated with the phase transition at zero temperature, is often generated by suppressing the magnetically ordered phases by applying hydrostatic pressure, magnetic field, and chemical substitution. Among the heavy-fermion compounds showing quantum critical behavior, CeMIn$_5$ (M=Co, Rh and Ir; HoCoGa$_5$-type tetragonal structure) is particularly interesting because it shows a variety of phenomena related to the antiferromagnetic (AF) spin fluctuations enhanced around the AF-QCP. CeRhIn$_5$ orders in an incommensurate AF phase with the modulation of $q=(1/2,1/2,0.297)$ below $T_{\rm N}=3.8\ {\rm K}$.\cite{rf:Hegger2000,rf:Bao2000} An itinerant helical (or spiral) spin-density-wave (SDW) state has been proposed to be the stable magnetic structure of this phase. It is revealed that the application of hydrostatic pressure weakly suppresses the AF phase, and then induces the superconducting phase ($T_{\rm c}\sim 2.3\ {\rm K}$) above $\sim 1-1.5\ {\rm GPa}$.\cite{rf:Hegger2000,rf:Mito2001,rf:Kawasaki2001,rf:Fisher2002,rf:Mito2003,rf:Llobet2004} On the other hand, CeCoIn$_5$ shows unconventional superconductivity below $T_{\rm c}=2.3\ {\rm K}$ at ambient pressure, characterized by an anomalously large specific heat jump ($\Delta C/\gamma T_{\rm c}=4.5$) at $T_{\rm c}$.\cite{rf:Petrovic2001} Furthermore, non-Fermi-liquid behavior is observed in the temperature variations of the electronic specific heat and the resistivity measured above the superconducting critical field $H_{\rm c2}$. It is considered that these features are ascribed to the AF quantum critical fluctuations: CeCoIn$_5$ is located near the AF-QCP.\cite{rf:Bianchi2003}

The nature of the ordered states in the solid solution CeRh$_{1-x}$Co$_x$In$_5$ is expected to provide crucial keys to clarifying the relationship between the incommensurate AF phase for CeRhIn$_5$ and the superconducting phase for CeCoIn$_5$. The specific heat, magnetic susceptibility and resistivity measurements for CeRh$_{1-x}$Co$_x$In$_5$\cite{rf:Zapf2001,rf:Jeffries2005} revealed that $T_{\rm N}$ of the incommensurate AF order for pure CeRhIn$_5$ is weakly reduced upon doping with Co, and then approaches zero at $x_{\rm c}\sim 0.7$. At the same time, the superconducting phase appears below $T_{\rm c}\sim 2\ {\rm K}$ in the $x$ range between $x=0.4$ and 1. The obtained $x-T$ phase diagram is quite similar to the pressure versus temperature ($P-T$) phase diagram for pure CeRhIn$_5$, implying that the superconducting phase in CeCoIn$_5$ closely correlates with the AF fluctuation enhanced at the boundary of the incommensurate AF phase. It is therefore interesting to investigate the magnetic properties in the intermediate $x$ range using microscopic probes. The unit-cell volume is found to shrink upon doping Co into CeRhIn$_5$. This is consistent with the difference in ionic radius between Rh and Co. In contrast, the tetragonal $c/a$ ratio is significantly enhanced with increasing $x$. We expect that the enhancement of the $c/a$ ratio may influence the AF structure in the intermediate $x$ range because the incommensurate axis for pure CeRhIn$_5$ is parallel to the $c$-axis. To clarify these points, we have examined the microscopic features of CeRh$_{1-x}$Co$_x$In$_5$ by performing elastic neutron scattering measurements. In this letter, we present the finding of a new magnetic phase evolving at $x=0.4$, where both the incommensurate AF and superconducting phases appear.

Single crystals of CeRh$_{0.6}$Co$_{0.4}$In$_5$ were grown by the In-flux method. Appropriate amounts of Ce, Rh, Co, and In as the flux were set in an alumina crucible, which was then sealed under 0.02 MPa Ar atmosphere in a quartz tube. The obtained ingots of single crystal, with a typical size of $3\times 15\times 10$ mm$^3$, were cut into several pieces by spark erosion. The specific heat $C_P$ was measured by the relaxation method in commercial heat-capacity equipment (Oxford Instruments) between 0.5 K and 100 K. The magnetic susceptibility $\chi$ was estimated from the magnetization values at $H=5\ {\rm kOe}$ in the temperature range of $2-360\ {\rm K}$, measured using a SQUID magnetometer (Quantum Design). Elastic neutron scattering experiments were performed using the triple-axis spectrometer GPTAS (4G) located at the research reactor, JRR-3M, of the JAEA, Tokai. We prepared a rod-type sample with the dimensions of $\sim 1.7\times 1.7\times 12\ {\rm mm}^3$ to minimize the effects of neutron absorption by Rh and In. The rod-type sample was mounted in a standard Al capsule filled with $^4$He gas so that the scattering plane was ($hhl$), and cooled to 1.4 K in a $^4$He cryostat. The neutron momentum $k=3.83\ {\rm \AA}^{-1}$ was selected using the (002) reflection of pyrolytic graphite (PG) for both the monochromator and the analyzer. For this momentum, the neutron penetration length in the sample was calculated to be $\sim 1.8\ {\rm mm}$. We chose a combination of 40'-40'-40'-80' collimators, and two PG filters to eliminate higher-order reflections.

Figure 1 shows the temperature variations of specific heat divided by temperature $C_P/T$. We observed two clear jumps in the $C_P/T$ data at $1.4\ {\rm K}$ ($\equiv T_{\rm c})$ and $3.5\ {\rm K}$ ($\equiv T_{\rm Nh}$), corresponding to the superconducting and incommensurate AF transitions, respectively. These transition temperatures are roughly in agreement with those reported previously ($T_{\rm c} \sim 1.5\ {\rm K}$ and $T_{\rm Nh}=3.6\ {\rm K}$).\cite{rf:Zapf2001} The $\Delta C_P/\gamma T_{\rm c}$ value (0.79) estimated from the jump of our $C_P/T$ data at $T_{\rm c}$ is also comparable to a previous result (0.7),\cite{rf:Zapf2001} but the jump at $T_{\rm c}$ in our $C_P/T_{\rm c}$ data is much sharper, probably due to a slight difference in the sample quality, or a deviation of $x$ from the starting composition (less than 8\%). It seems that the jump anomaly in $C_P/T$ at $T_{\rm c}$ is not entropy-balanced. Our preliminary experiments on the specific heat for magnetic fields greater than $H_{\rm c2} \sim 70\ {\rm kOe}$ revealed that $C_P/T$ unusually increases below $T \sim 1.5\ {\rm K}$, indicating that some kind of fluctuation still exists in the low-temperature region. On the other hand, we have found a clear kink anomaly in $C_P/T$ at $2.9\ {\rm K}\ (\equiv T_{\rm Nc})$. This anomaly can be ascribed to the commensurate AF transition from the present neutron scattering experiments, the details of which are described later. We have estimated the entropy release associated with the phase transitions by simply using the relation $S=\int_{0.5{\rm K}}^{T}C_P/T\,dT$. The $S$ value at $T_{\rm Nh}$ is estimated to be $\sim 0.38R\ln2$, which is much smaller than the $R\ln2$ expected from the crystalline-electric-field doublet ground state of the Ce 4f electron under tetragonal symmetry. It is suggested that the small entropy value at $T_{\rm Nh}$ and the enhancement of $C_P/T$ above $T_{\rm Nh}$ can be attributed to many-body effects, such as the Kondo effect and short-range magnetic correlations. In contrast to the clear anomalies seen in $C_P/T$ at $T_{\rm Nh}$ and $T_{\rm Nc}$, the phase transitions at $T_{\rm Nh}$ and $T_{\rm Nc}$ are found to weakly affect the $a$-axis magnetic susceptibility $\chi_a(T)$ (inset of Fig.\ 1). A broad maximum is observed in $\chi_a(T)$ at $\sim 5.5\ {\rm K}$, where $C_P/T$ begins to develop. We have found no clear anomaly at $T_{\rm Nh}$ in $\chi_a(T)$, while a weak inflection point at $T_{\rm Nc}$.
\begin{figure}[tbp]
\begin{center}
\includegraphics[keepaspectratio,width=0.4\textwidth]{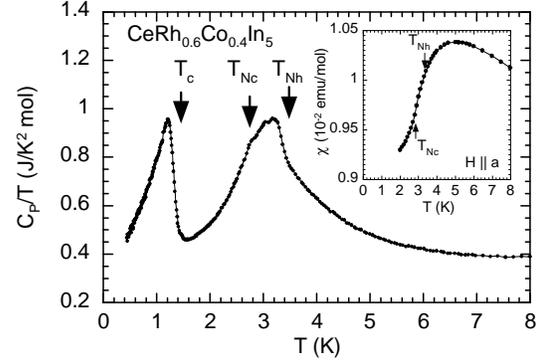}
\end{center}
  \caption{Temperature variations of specific heat divided by temperature $C_P/T$ for CeRh$_{0.6}$Co$_{0.4}$In$_5$. The inset shows temperature variations of the $a$-axis magnetic susceptibility $\chi_a$ for CeRh$_{0.6}$Co$_{0.4}$In$_5$, obtained at a magnetic field of $H=5\ {\rm kOe}$.
}
\end{figure}

Displayed in Fig.\ 2 is the neutron scattering pattern obtained from scans at the momentum transfers $Q=(1/2,1/2,1+\xi)$ ($0\le \xi \le 1$) at 1.4 K, where the instrumental background and higher-order nuclear diffractions were carefully subtracted using the data at 5 K. We have found that three nonequivalent Bragg peaks, other than the nuclear Bragg peaks expected for the tetragonal structure, develop at low temperatures. The peaks are observed at the corresponding $Q$ positions in the different crystalline Brillouin zones investigated. Taking appropriate averages over all peaks for each equivalent $Q$ position, we obtained the three wave vectors that characterize the observed Bragg peaks as $q_{\rm h} = (1/2, 1/2, 0.306(10)(\equiv \delta))$, $q_1 = (1/2, 1/2, 0.402(12))$ and $q_{\rm c} = (1/2, 1/2, 1/2)$. Note that the wave vector (1/2, 1/2, 0.598), which is equivalent to $q_1$, can be described as $(1/2, 1/2, 2\delta)$ within experimental error. This strongly suggests that $q_1$ corresponds to the second higher harmonic of $q_{\rm h}$. To emphasize this possibility, we choose $q_{\rm 2h} = (1/2, 1/2, 2\delta)$ instead of $q_1$ throughout this article as the expression of relevant wave vectors, together with $q_{\rm h}$ and $q_{\rm c}$. 
\begin{figure}[bp]
\begin{center}
\includegraphics[keepaspectratio,width=0.4\textwidth]{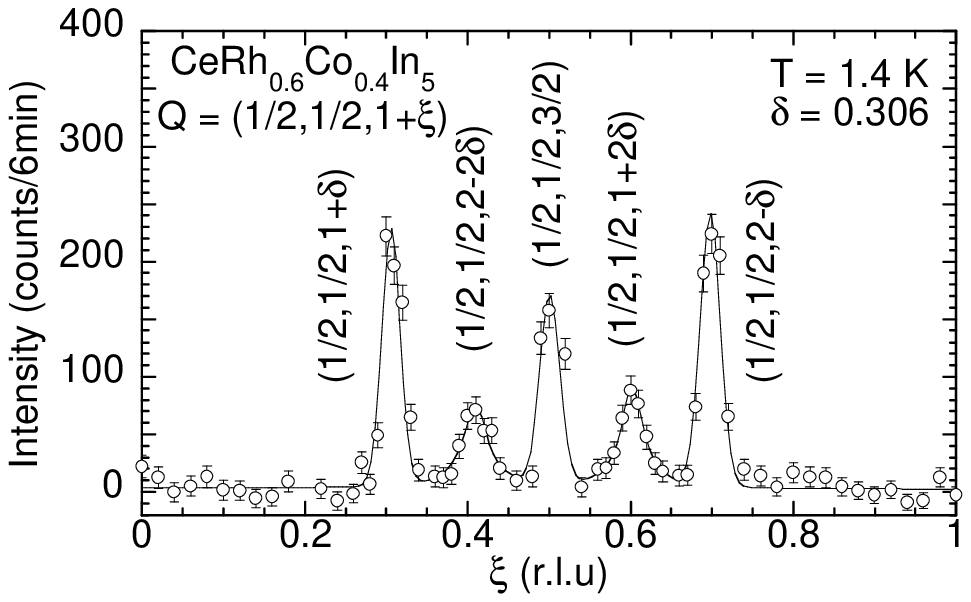}
\end{center}
  \caption{Neutron scattering pattern obtained by scans at $Q=(1/2,1/2,1+\xi)$ ($0\le \xi \le 1$) at 1.4 K for CeRh$_{0.6}$Co$_{0.4}$In$_5$. The line is a guide for the eye.
}
\end{figure}

No additional Bragg peak is found in the scan at $Q=(0,0,1+\xi)$ with $0\le \xi\le 1$. It is considered that the appearance of the Bragg peaks for $q_{\rm h}$, which correspond to the peaks at $Q=(1/2,1/2,1+\delta)$ and ($1/2,1/2,2-\delta$) in Fig.\ 2, can be attributed to the incommensurate AF order, since pure CeRhIn$_5$ also shows this type of order with $q=(1/2,1/2,0.297)$.\cite{rf:Bao2000} Although the error in $\delta$ in the present experiments is somewhat large, the obtained $\delta$ value coincides fairly well with that for pure CeRhIn$_5$, indicating that the period of the incommensurate AF structure is insensitive to Co doping. A similar feature is also observed in pure CeRhIn$_5$ under hydrostatic pressure.\cite{rf:Llobet2004}

The development of new Bragg peaks relevant to the wave vectors of $q_{\rm c}$ and $q_{\rm 2h}$ indicates that new orders evolve with intermediate Co concentrations. The intensity of the Bragg peaks for $q_{\rm c}$ is found to show a tendency to decrease with increasing $|Q|$, strongly suggesting that they originate from the magnetic order, i.e., the commensurate AF order with the modulation of $q_{\rm c}$. On the other hand, the positions of the Bragg peaks for $q_{\rm 2h}$ match considerably the places where the harmonics of the incommensurate AF modulation are expected to appear, implying that these Bragg peaks are tightly coupled with the incommensurate AF order. However, the widths of these Bragg peaks are larger than those for the incommensurate AF order (and also for the commensurate AF order) by a factor of $\sim 1.2$. We thus consider that the order with the modulation of $q_{\rm 2h}$ occurs in a shorter range along the $c$-axis than the incommensurate AF order.

We have calculated the magnetic scattering amplitudes $\mu f(|Q|)$ for the incommensurate and commensurate AF orders from the integrated intensities of the magnetic Bragg peaks, obtained from longitudinal and transverse scans. In the present experiments, we cannot determine whether the incommensurate AF phase has a helical or a sinusoidal SDW structure. We here assume a helical SDW structure in accordance with a previous report.\cite{rf:Bao2000} The present experimental results are basically unaffected even if the incommensurate AF phase has a sinusoidal SDW structure. We have assumed the directions of both the helical and commensurate AF moments to be along the tetragonal basal plane. The former AF structure is the same as that observed in pure CeRhIn$_5$.\cite{rf:Bao2000} The helical structure is known to involve the internal degrees of freedom corresponding to a spiral orientation. However, they usually do not affect the data obtained by the unpolarized neutron scattering technique. We thus simply calculated the $\mu f(|Q|)$ values for the helical AF order without considering such an effect. On the other hand, it is expected that the commensurate AF structure with the modulation of $q_{\rm c}$ and the moment along the tetragonal basal plane makes two equivalent magnetic domains. We here estimated the $\mu f(|Q|)$ value for the commensurate AF order by assuming these domains to be equally distributed. In such a case, we cannot determine the direction of the moment in the tetragonal basal plane. In Fig.\ 3, we plot the $|Q|$ variations of the magnetic form factors $f(|Q|)$ at 1.4 K for the helical and commensurate AF orders. The obtained $f(|Q|)$ curves for both AF orders show monotonic decreases with increasing $|Q|$, supporting our assumptions on the AF structures. In addition, they are in good agreement with the $f(|Q|)$ curve calculated for Ce$^{3+}$ ion,\cite{rf:IntTable} indicating that both the AF orders are mainly formed by Ce 4f moments. We also attempted to calculate the $f(|Q|)$ values by assuming the moment in the commensurate AF order to be along the $[111]$ direction, as a promising candidate for the AF structure. However, the obtained $f(|Q|)$ curve shows an improperly large oscillation due to the existence of the $c$-component of the ordered moment. We thus exclude such a possibility for the commensurate AF structure.
\begin{figure}[tbp]
\begin{center}
\includegraphics[keepaspectratio,width=0.4\textwidth]{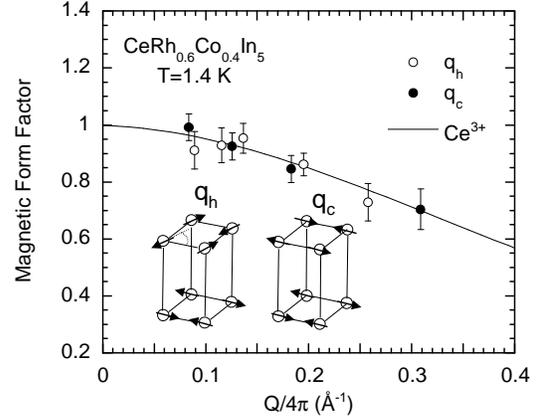}
\end{center}
  \caption{Magnetic form factor $f(|Q|)$ at 1.4 K for incommensurate and commensurate AF orders with modulations of $q_{\rm h}=(1/2,1/2,\delta)$ ($\delta = 0.306$) and $q_{\rm c}=(1/2,1/2,1/2)$, respectively. The $f(|Q|)$ values for the incommensurate AF order are estimated by assuming a helical SDW structure. The $f(|Q|)$ function calculated for Ce$^{3+}$ ion\cite{rf:IntTable} is also shown. The inset shows the arrangements of the magnetic moment on the Ce ions for helical and commensurate AF structures.
}
\end{figure}

We have derived the magnitudes of the volume-averaged AF moments for the incommensurate AF phase ($\mu_{\rm h}$) and commensurate AF phase ($\mu_{\rm c}$) from the $\mu f(0)$ values, using the integrated intensities of the weak (110) nuclear Bragg peak as a reference. $\mu_{\rm h}$ and $\mu_{\rm c}$ at 1.4 K are estimated to be 0.38(3) $\mu_{\rm B}/{\rm Ce}$ and 0.21(2) $\mu_{\rm B}/{\rm Ce}$, respectively, which are much smaller than the magnitude (0.75 $\mu_{\rm B}/{\rm Ce}$) of the helical AF moment for pure CeRhIn$_5$. Furthermore, the magnitude of the total AF moment $\mu_{\rm T}=\sqrt{\mu_{\rm h}^2+\mu_{\rm c}^2}=0.43(5)\ \mu_{\rm B}/{\rm Ce}$ is also found to be smaller, indicating that the ordered moments are strongly reduced in the intermediate $x$ range.

In Fig.\ 4, we show the temperature variations of the integrated intensities $I(T)$ of the Bragg peaks for $Q=(1/2,1/2,1-\delta)$, $(1/2,1/2,1-2\delta)$, and $(1/2,1/2,3/2)$. We observed that the Bragg-peak intensity for the incommensurate AF order, $I_{\rm h}(T)$, starts increasing at $\sim 3.5\ {\rm K}$, and then becomes nearly constant below $\sim 2.3\ {\rm K}$ (Fig.\ 4(a)). Since the onset of $I_{\rm h}(T)$ coincides with $T_{\rm Nh}$, we consider that the jump observed in $C_P/T$ at $T_{\rm Nh}$ is attributed to the incommensurate AF order. Interestingly, the Bragg-peak intensity for the $q_{\rm 2h}$-order $I_{\rm 2h}(T)$ also develops below $T_{\rm Nh}$ (Fig.\ 4(b)), strongly suggesting that the $q_{\rm 2h}$ order accompanies the evolution of the incommensurate AF order. On the other hand, we found that the Bragg-peak intensity for the commensurate AF order $I_{\rm c}(T)$ develops at a clearly lower temperature than $T_{\rm Nh}$ (Fig.\ 4(c)). The onset of $I_{\rm c}(T)$ is estimated to be $\sim 2.9\ {\rm K}$, which corresponds to $T_{\rm Nc}$ defined by the kink in the $C_P/T$ data. This correspondence indicates that the commensurate AF phase appears as the static order. $I_{\rm c}(T)$ linearly develops below $T_{\rm Nc}$, and shows a tendency to saturate below $\sim 1.7\ {\rm K}$. Such behavior of the $I_{\rm c}(T)$ curve differs greatly from that expected from the simple AF order, and may be caused by an effect of correlations between the different types of AF order. We suggest that the commensurate AF order does not replace the incommensurate AF order but coexists with it in the sample, since $I_{\rm h}(T)$ is not significantly reduced below $T_{\rm Nc}$. It is not clear in the present experiments whether these AF phases are microscopically coexistent or inhomogeneously separated. This issue may be resolved by performing NMR and $\mu$SR experiments in the future.
\begin{figure}[tbp]
\begin{center}
\includegraphics[keepaspectratio,width=0.4\textwidth]{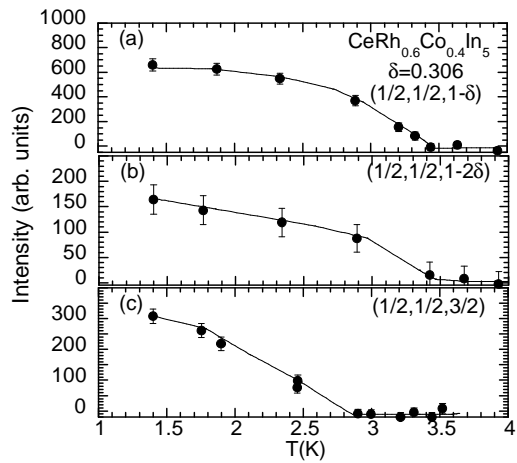}
\end{center}
  \caption{Temperature variations of Bragg-peak intensities for (a) $Q=(1/2,1/2,1-\delta)$, (b) $Q=(1/2,1/2,1-2\delta)$ and (c) $Q=(1/2,1/2,3/2)$ with $\delta = 0.306$, which evolve owing to incommensurate AF order, order with the modulation of $q_{\rm 2h}$, and commensurate AF order, respectively. The lines are guides for the eye.
}
\end{figure}

We found the evolution of the commensurate AF phase with the wave vector of $q_{\rm c}$ in  CeRh$_{0.6}$Co$_{0.4}$In$_5$. This modulation is the same as that for the AF phase observed in CeIn$_3$.\cite{rf:Walker97,rf:Morin88} The crystal structure of CeRhIn$_5$ is closely correlated with that of CeIn$_3$. CeRhIn$_5$ has a tetragonal structure with the layers of CeIn$_3$ and RhIn$_2$ stacked sequentially along the $c$ axis, while CeIn$_3$ crystallizes in the AuCu$_3$-type cubic structure. On the other hand, it has been very recently revealed that the commensurate AF phase with the same structure as that observed in CeRh$_{0.6}$Co$_{0.4}$In$_5$ also evolves in the Ir-doped CeRhIn$_5$ system.\cite{rf:Christianson2005} In CeRh$_{1-x}$Ir$_{x}$In$_5$, the incommensurate AF phase is weakly suppressed by increasing $x$, and then vanishes at $x\sim 0.6$.\cite{rf:Pagliuso2001,rf:Kawasaki2006} At the same time, the superconducting phase develops above $x\sim 0.3$. The new commensurate AF phase is found to appear in the $x$ range between 0.3 to 0.6. This $x-T$ phase diagram is quite similar to that for CeRh$_{1-x}$Co$_{x}$In$_5$. We thus expect that the commensurate magnetic correlation with the modulation of $q_{\rm c}$ commonly exists in CeIn$_3$ and CeRhIn$_5$, and increases upon doping Co or Ir into CeRhIn$_5$. Since these alloys show superconductivity near the commensurate AF phase, we suggest that the commensurate magnetic correlation is coupled with the evolution of superconductivity. 

We consider that the enhancement of the commensurate AF correlation may be governed by the tetragonal $c/a$ ratio. The $c/a$ ratio is enhanced by doping with Co, but reduced by doping with Ir.\cite{rf:Zapf2001} We expect that the $c/a$ ratio is tightly connected with the shapes of the Fermi surfaces, and therefore these variations influence the magnitude of the incommensurate magnetic correlation more sensitively than the commensurate one. If the incommensurate magnetic correlation is reduced by changing the $c/a$ ratio from that for pure CeRhIn$_5$, the commensurate AF correlation becomes relatively large, generating the commensurate AF order.

In the present experiments, we have also found the existence of Bragg peaks with the modulation of $q_{\rm 2h}$. It is natural to think that these peaks are attributed to the second-order harmonics of the incommensurate AF order. One of the probable origins of these Bragg peaks is the strain-wave order accompanying the helical or sinusoidal SDW order through the strong spin-orbit interactions and the magneto-elastic couplings. It is well known that the SDW orders in Cr are accompanied by the formation of a strain wave.\cite{rf:Fawcett94} Elastic neutron scattering experiments for Cr revealed that the amplitude of the strain wave is proportional to the square of the SDW moment,\cite{rf:Pynn76} which is consistent with theory.\cite{rf:Kotani76} In neutron scattering experiments, the ratio of the Bragg-peak intensities of the primary SDW and the strain wave, $I_{\rm SW}/I_{\rm SDW}$, is usually expected to be of the order of $10^{-2}-10^{-3}$, which is much smaller than the ratio of the intensities $I_{\rm 2h}/I_{\rm h}\sim 0.3$ obtained from the present measurements. On the other hand, we cannot exclude the possibility that another AF order with the modulation of $q_1$ occurs in parts of the sample. The origin of these Bragg peaks with large widths and their relationship to the incommensurate AF order are unclear at the present stage. We are now planning to perform polarized neutron scattering and X-ray scattering experiments to clarify these points.

In summary, our elastic neutron scattering experiments for the mixed compound CeRh$_{0.6}$Co$_{0.4}$In$_5$ revealed that it shows not only the incommensurate AF and superconducting orders below $T_{\rm Nh}=3.5\ {\rm K}$ and $T_{\rm c}=1.4\ {\rm K}$, respectively, but also the commensurate AF order below $T_{\rm Nc}=2.9\ {\rm K}$. Since this commensurate AF phase is commonly observed near the superconducting phase of Co- or Ir-doped CeRhIn$_5$,\cite{rf:Christianson2005} we suggest that the commensurate AF and the superconducting phases are coupled to each other. We also found the Bragg peaks that correspond to the second-order harmonic of the incommensurate AF order.

\section*{Acknowledgements}
One of us (M.Y.) is grateful to J.\ Igarashi and Y.\ Tabata for helpful discussions and comments. This work was supported by a Grant-in-Aid for Scientific Research from the Ministry of Education, Culture, Sports, Science and Technology.

\end{document}